# Nonlinear Photoluminescence Properties of Trions in Hole-doped Single-walled Carbon Nanotubes

Naoto Akizuki[1], Munechiyo Iwamura[1], Shinichiro Mouri[1], Yuhei Miyauchi[1,2],

Tomohiro Kawasaki[3], Hiroshi Watanabe[3], Tohru Suemoto[3], Kouta Watanabe[4],

Kenichi Asano[4], and Kazunari Matsuda[1,+]

[1]*Institute of Advanced Energy, Kyoto University, Uji, Kyoto 611-0011, Japan*

[2]*Japan Science and Technology Agency, PRESTO, 4-1-8 Honcho Kawaguchi, Saitama 332-0012, Japan*

[3]*Institute of Solid State Physics, University of Tokyo, 5-1-5 Kashiwanoha, Kashiwa-shi, Chiba 277-8581, Japan*

[4]*Department of Physics, Osaka University, Toyonaka, Osaka 560-0043, Japan*

## ABSTRACT

We studied the excitation density dependence of photoluminescence (PL) spectra of excitons and trions (charged excitons) in hole-doped single-walled carbon nanotubes. We found that the PL intensity of trions exhibited a strong nonlinear saturation behavior as the excitation density increased, whereas that of excitons exhibited a weak sublinear behavior. The strong PL saturation of trions is attributed to depletion of doped holes that are captured by excitons in the formation processes. Moreover, the effective radiative lifetime of a trion was evaluated to be approximately 20 ns.

Single-walled carbon nanotubes (SWNTs) are a type of nanomaterial that has engendered substantial current interest from various viewpoints such as their mechanical, electronic, and optical properties [1]. SWNTs with a cylindrical structure having a diameter of approximately 1 nm have nearly ideal quasi one-dimensional (1D) electronic states [2-4], and Coulomb interactions between the optically generated electron-hole pairs in 1D SWNTs result in the formation of strongly bound exciton states [5]. Stable excitons, which persist even at room temperature, play an important role in the optical properties of SWNTs [6-8]. Recently, it has been reported that optically generated excitons and doped holes form trions (charged excitons) even at room temperature in carrier-doped SWNTs [9,10], and trions in carrier-doped SWNTs have been extensively studied because of their importance [11-16]. Because of their non-zero charge and spin degrees of freedom, trions are expected to be a fundamental element facilitating charge and spin manipulation in semiconductors. Stable trions in SWNTs are, thus, expected to have superior characteristics toward spin manipulation at room temperature, and a longer spin coherence time due to weak spin-orbit interaction relative to trions in other semiconductor systems.

An understanding of the nonlinear optical properties of low-dimensional semiconductors provides insightful information concerning their fundamental physics and potential applications. Strong absorption saturation due to filling of the exciton state has been observed for non-doped SWNTs [17], and this phenomenon has been utilized in the application of SWNTs as saturable absorbers for mode-locked fiber lasers [18,19]. The nonlinear saturation behavior of exciton photoluminescence (PL) under strong excitation conditions has also been extensively studied [20-23]. The origin of this strong nonlinear PL behavior has been clarified as exciton-exciton annihilation (EEA) processes [21,24-27] originating from rapid 1D migration [28] and collision of excitons in SWNTs. A comparison of the nonlinear PL behavior of trions with those of excitons may provide important information concerning exciton and trion dynamics, which could also lead to a further understanding of the internal spin and valley configurations of trions that contribute to PL in carrier-doped SWNTs.

In this study, we investigated the excitation density dependence of PL spectra of excitons and trions in hole-doped SWNTs. We found that the PL intensity of trions in hole-doped SWNTs exhibits a strong nonlinear saturation behavior as the excitation density increases, whereas that of excitons demonstrates a weak sublinear behavior. The strong PL saturation of trions is attributed to depletion of doped holes that are captured by excitons in the formation processes of trions. We also discuss the internal spin and valley configurations of a trion from the experimentally deduced effective radiative decay rate.

To prepare hole-doped SWNTs [9], 0.14 wt% CoMoCAT SWNTs were dispersed in a toluene solution containing 0.14 wt% poly[9,9-dioctylfluorenyl-2,7-diyl] (PFO) [29] by bath sonication for 1 h. Then, vigorous sonication using a high-power tip-type ultrasonic homogenizer was performed for 20 h, and ultracentrifugation was performed at 31000 g for 20 min. For hole-doping of the SWNT solution, 2,3,5,6-Tetrafluoro-7,7,8,8-tetracyanoquinodimethane ($F_4$TCNQ) dispersed in toluene was added to PFO-dispersed SWNTs as a *p*-type dopant [30]. The PL quantum yield $\eta$ was determined using the reference dye, Styryl 13 with $\eta$(Styryl 13) = 3%. The excitation density dependence of PL spectra of non-doped and hole-doped SWNTs was measured using a Ti:Al$_2$O$_3$ pulsed laser at a variable excitation density of 0.01–1.2 mJ/cm$^2$. The excitation photon energy and repetition rate of the pulses were 1.55 eV and 80 MHz, respectively. A typical pulse width of the laser was estimated as approximately 150 fs. The PL signals were detected through a monochromator equipped with a liquid-nitrogen cooled InGaAs photodiode array.

Figure 1 shows typical PL spectra of (7,5) SWNTs hole-doped with various concentrations of F$_4$TCNQ molecules and excited at 1.55 eV. The PL intensity of the $E_{11}$ exciton transition (X) at around 1.19 eV drastically decreases because of hole-doping of SWNTs which is consistent with previously reported results [30]. The PL peak around 1.01 eV arising from the recombination of trions (X$^+$) because of hole-doping [9] is observed in Fig. 1. The inset shows the PL spectra of non-doped SWNTs with excitation photon energy of 1.55 eV, as a reference. For non-doped SWNTs, the PL quantum yield of the $E_{11}$ exciton (X) peak $\eta_{\text{non–doped}}(X)$ was evaluated to be ~1%, which is consistent with a previously reported value [29]. For SWNTs hole-doped with a F$_4$TCNQ concentration of 300 μg/ml, the PL quantum yields of the $E_{11}$ exciton (X) peak $\eta_{\text{doped}}(X)$ and the trion (X$^+$) peak $\eta_{\text{doped}}(X^+)$ were evaluated to be ~0.01% and ~0.01%, respectively [32]. The PL quantum yields of the $E_{11}$ exciton (X) and trion (X$^+$) peaks in hole-doped SWNTs are much smaller than that of the $E_{11}$ exciton (X) peak in non-doped SWNTs, which will be discussed below.

We examined the excitation density dependence of PL spectra of non-doped and hole-doped SWNTs. Figure 2(a) shows the PL spectra of non-doped SWNTs with various excitation densities of 0.01–1.2 mJ/cm$^2$. The PL spectra are normalized by the corresponding excitation densities. The normalized PL intensity of the $E_{11}$ exciton (X) peak gradually decreases with increasing excitation density suggesting saturation of the PL intensity of the $E_{11}$ exciton (X) peak at higher excitation conditions. Figure 2(b) shows the PL spectra of SWNTs hole-doped with a F$_4$TCNQ concentration of 300 μg/ml normalized by the corresponding excitation densities. The normalized PL intensities of the $E_{11}$ exciton (X) and trion (X$^+$) peaks gradually decrease with increasing excitation density. The relative intensities between the $E_{11}$ exciton (X) and trion (X$^+$) peaks change suggesting that the saturation behavior of the PL intensity is different between the $E_{11}$ exciton (X) and trion (X$^+$) peaks.

We plotted the integrated PL intensity of the $E_{11}$ exciton (X) peak of non-doped SWNTs as a function of excitation density in Fig. 3(a). The dotted line represents a linear plot normalized by the experimental data around the weak excitation density regime. The integrated PL intensity of the $E_{11}$ exciton (X) peak linearly increases in the lower excitation density regime and gradually saturates in the higher excitation density regime. The saturation behavior of the $E_{11}$ exciton (X) peak in the higher excitation density regime, where many excitons are generated in the SWNT, can be understood as a consequence of the opening of non-radiative recombination paths due to exciton-exciton annihilation (EEA) processes [21].

We also plotted the excitation density dependence of the integrated PL intensity of the $E_{11}$ exciton (X) peak (solid circles) and trion (X$^+$) peak (open circles) in SWNTs hole-doped with a F$_4$TCNQ concentration of 300 μg/ml in Fig. 3(b). The dotted lines represent linear plots normalized by the experimental data in the weak excitation density regime. Figure 3(b) shows that the integrated PL intensities of the $E_{11}$ exciton (X) and trion (X$^+$) peaks are proportional to the excitation density in the weak excitation regime and gradually saturate in the strong excitation regime. However, the saturation behaviors of the two peaks are different, that is, the integrated PL intensity of the trion (X$^+$) peak shows a stronger saturation behavior than that of the $E_{11}$ exciton (X) peak. These nonlinear behaviors do not arise from heating the samples by laser excitation (see Ref. 32). Moreover, the nonlinear behavior of the integrated PL intensity of the $E_{11}$ exciton (X) peak in hole-doped SWNTs becomes weaker in comparison with that of the $E_{11}$ exciton (X) peak in non-doped SWNTs, as shown in Fig. 3(a).

We next examine the dynamics of excitons and trions in hole-doped SWNTs to understand the observed nonlinear PL behavior. The rate equation model is used in the analysis within the framework of a three-level system consisting of exciton (X), trion (X$^+$), and ground (gr) states, as shown in the inset of Fig. 3(b). Here we considered the EEA processes for excitons, formation processes of trions from excitons by capture of doped holes, and the filling of trion states. The populations of excitons $N_\text{X}(t)$ and trions $N_{\text{X}^+}(t)$ at time $t$ are expressed as

$$\frac{dN_\text{X}(t)}{dt} = G(t) - \gamma_\text{X} N_\text{X}(t) - \gamma_\text{EEA} N_\text{X}(t)^2 - \gamma_\text{h}[Q - N_{\text{X}^+}(t)]N_\text{X}(t), \quad (1)$$

$$\frac{dN_{X^+}(t)}{dt} = \gamma_h [Q - N_{X^+}(t)] N_X(t) - \gamma_{X^+} N_{X^+}(t), \qquad (2)$$

where $G(t)$ is the generation function of excitons approximated by a Gaussian pulse and $Q$ (~0.18 hole/nm, doped with a F$_4$TCNQ concentration of 300 μg/ml) is the number density of doped holes in a hole-doped SWNT derived from the decrease of oscillator strength (bleaching) in absorption spectra [31]. The variables $\gamma_X, \gamma_{X^+}, \gamma_{EEA}, \gamma_h$ in Eqs. (1) and (2) are the decay rates of excitons and trions, respectively, to the ground state, EEA rate coefficient, and formation rate coefficient of trions from excitons by the capture of doped holes, respectively. For analysis of the rate equations in comparison with the experimental data, the integrated population of excitons (trions) $N_X(N_{X^+})$ is calculated as $N_{X(X^+)} = \int_0^\infty N_{X(X^+)}(t)dt$. We also used the relationship between the integrated PL intensity of excitons (trions) $I_X(I_{X^+})$ and population of excitons (trions), $I_{X(X^+)} \propto \eta(X(X^+))N_{X(X^+)}$.

Under weak excitation density conditions for excitons (X) in non-doped SWNTs, the decay rate $\gamma_X$ in Eq. (1) is evaluated to be (~30 ps)$^{-1}$ from the relationship $\gamma_X = \gamma_{Rad}(X)/\eta_{non-doped}(X)$ where $\gamma_{Rad}(X)$ is (~3 ns)$^{-1}$ [33] and $\eta_{non-doped}(X)$ is ~1%. For excitons in SWNTs hole-doped with a F$_4$TCNQ concentration of 300 μg/ml, the PL decay rate is expressed as $\gamma_X + \gamma_h[Q - N_{X^+}(t)] \sim \gamma_{Rad}(X)/\eta_{doped}(X)$ and evaluated to be (~450 fs)$^{-1}$, where we considered $\eta_{doped}(X)$ to be ~0.01% and that the change of $\gamma_{Rad}(X)$ depended on the doped hole density [31]. Since the PL decay rate of excitons in non-doped SWNTs $\gamma_X$ is much smaller than that in hole-doped SWNTs, the condition $\gamma_h [Q - N_{X^+}(t)] \gg \gamma_X$ is fulfilled for hole-doped SWNTs and $\gamma_h Q$ is taken to be $(\sim 450 \text{ fs})^{-1}$ by considering that $Q \gg N_{X^+}(t)$ in weak excitation condition.

We solved the rate equations given by Eqs. (1) and (2) for populations of excitons and trions using the parameters deduced above ($\gamma_X, \gamma_h, Q$) and by determining $\gamma_{EEA}$ and $\gamma_{X^+}$ as fitting parameters. The experimentally observed nonlinear behavior of the integrated PL intensity of the exciton (X) and trion (X$^+$) peaks are well reproduced both by the calculated solid black line for excitons (X) and solid red line for trions (X$^+$) in Fig. 3(b). The fitted parameter of the EEA rate coefficient $\gamma_{EEA}$ of $(\sim 20 \text{ ps})^{-1}$ for a nanotube having a length $L$ of 0.3 μm is smaller than previously reported values for non-doped SWNTs [22,24] (see Ref. 32). The determined decay rate of trions $\gamma_{X^+}$ of $(\sim 800 \text{ fs})^{-1}$ is nearly consistent with the reported value [15] and the experimentally measured value [32].

Here, we discuss the mechanisms of the nonlinear PL behavior of excitons and trions in hole-doped SWNTs. The nonlinear PL behavior of the exciton (X) peak of hole-doped SWNTs is weaker than that of non-doped SWNTs, as shown in Figs. 3(a) and 3(b), which can be explained as follows. Very fast exciton relaxation (caused by trion formation from an exciton) of the order of $\gamma_h Q$ $((\sim 450 \text{ fs})^{-1})$ relative to that of EEA processes $((20 \text{ ps})^{-1})$ occurs predominantly, which weakens the nonlinear PL behavior of excitons caused by the EEA processes in hole-doped SWNTs. In contrast, the strong nonlinear PL behavior of the trion (X$^+$) peak comes from depletion of doped holes that are captured by excitons to form trions, owing to their limited number ($Q \cdot L$ is ~50 holes for a F$_4$TCNQ concentration of 300 µg/ml) in a chemically hole-doped SWNT.

The nonlinear PL behaviors of trions (X$^+$) in hole-doped SWNTs with various F$_4$TCNQ concentrations were measured to confirm the validity of the mechanism described above. Figure 4 shows the excitation density dependence of the integrated PL intensity of the trion (X$^+$) peak of SWNTs hole-doped with various F$_4$TCNQ concentrations. The experimental curves are normalized around the weak excitation density regime, and the dotted line shows the linear behavior. Here, we define the PL saturation density, where the experimental data deviates from the dotted linear line, shown as an arrow for a F$_4$TCNQ concentration of 300 µg/ml in Fig. 4. The PL saturation density, as shown in Fig. 4, gradually shifts to the higher excitation regime with increasing doped hole concentration. This experimental result can be easily explained by the mechanism described above. Because many optically generated trions are needed for the full occupation of doped holes in heavily hole-doped SWNTs, the PL saturation density shifts to the higher excitation regime in a manner that strongly depends on the concentration of doped holes.

Here, we discuss the effective radiative decay rate of a trion, which is expressed as $\gamma_{\text{Rad}}(X^+) \sim \gamma_{X^+} \cdot \eta_{\text{doped}}(X^+) \cdot A$, where $A$ corresponds to the distribution factor arising from the degeneracy of states. The spin and valley configurations of the lowest optically allowed trion states are ($N_c^K$, $S$) = (0, 1/2) or (−1, 1/2), where $N_c^K \equiv \sum(e_\alpha^{K\dagger} e_\alpha^K - h_\alpha^{K\dagger} h_\alpha^K)$ is the number of charges in the K valley, $e_\alpha^K$ and $h_\alpha^K$ are the annihilation operators of the electron and hole, respectively, and $S$ is the total electron and hole spin. Their degeneracies are both given by $g = 2$, which are nothing but the spin degeneracies [12]. Whereas, the lowest optically forbidden trion states with ($N_c^K$,$S$) = (0,3/2) and (−1,3/2) have the energy almost identical to that of the lowest optically allowed trion. Their degeneracies are both given by $g' = 4$. Consequently, the distribution factor is estimated as $A \equiv g/(g+g') = 1/3$ [33]. Thus, the effective radiative decay rate of an optically allowed trion states with ($N_c^K$,$S$) = (0,1/2) and (−1,1/2) is evaluated approximately as $\gamma_{Rad}(X^+)$ ~(20 ns)$^{-1}$ from the experimental results of $\eta_{doped}(X^+)$ (~0.01%), the obtained decay rate of a trion $\gamma_{PL}(X^+)$ ((~800 fs)$^{-1}$) and $A$ (=1/3). The estimated radiative decay rate for a trion of (20 ns)$^{-1}$ is smaller than that for a singlet $E_{11}$ exciton of (3 ns)$^{-1}$ [33].

In summary, we studied the nonlinear PL behavior of excitons and trions in hole-doped SWNTs. The excitation density dependence of the trion PL intensity is significantly different from that of the exciton, and the trion PL intensity shows strong saturation behavior in contrast to the weak saturation behavior exhibited by the exciton PL intensity. This strong saturation of trion PL intensity was attributed to depletion of doped holes that are captured by excitons to form trions. Moreover, the effective radiative decay rate of a trion was deduced to be (20 ns)$^{-1}$. These insights into trion dynamics and spin configuration provide us with important evidence for obtaining spin manipulation in photoexcited SWNTs.

# Acknowledgement

This study was supported by a Grant-in-Aid for Scientific Research from the Japan Society for Promotion of Science (Grants 22740195, 22016007, 23340085, and 24681031), the Precursory Research for Embryonic Science and Technology program from the Japan Science and Technology Agency, the Asahi Glass Foundation, and the Yamada Science Foundation.

**FIGURE CAPTIONS**

**Figure 1** PL spectra of (7,5) SWNTs hole-doped with various F$_4$TCNQ concentrations. The (7,5) $E_{11}$ exciton and trion peaks are denoted as X and X$^+$, respectively. The excitation photon energy is 1.55 eV. The inset shows the PL spectrum of non-doped (7,5) SWNTs with an excitation photon energy of 1.55 eV.

**Figure 2** (a) Excitation density dependence of the PL spectra of non-doped (7,5) SWNTs. The $E_{11}$ exciton peak is denoted as X. (b) Excitation density dependence of the PL spectra of (7,5) SWNTs hole-doped with a F$_4$TCNQ concentration of 300 µg/ml. The $E_{11}$ exciton and trion peaks are denoted as X and X$^+$, respectively.

**Figure 3** (a) Excitation density dependence of the integrated PL intensity of the $E_{11}$ exciton (X) peak (1.19 eV, solid circles) of non-doped SWNTs. The dotted line represents the linear behavior normalized with experimental data around the weak excitation density conditions. (b) Excitation density dependence of the integrated PL intensity of the $E_{11}$ exciton (X) peak (1.19 eV, black solid circles) and the trion (X$^+$) peak (1.01 eV, red solid circles) of SWNTs hole-doped with F$_4$TCNQ concentration of 300 µg/ml. The solid circles, solid curves, and dotted lines represent the experimental data, the fitting of the experimental data to the rate equations, and the linear behavior normalized with experimental data around the weak excitation density conditions, respectively. The data points and fitted lines of the $E_{11}$ exciton (X) and trion (X$^+$) peaks are denoted as black and red, respectively. The inset shows a schematic of the three-level system of hole-doped SWNTs consisting of the exciton (X), trion ($X^+$), and ground (gr) states.

**Figure 4** Excitation density dependence of the integrated PL intensity of the trion ($X^+$) peak of SWNTs hole-doped with various $F_4$TCNQ concentrations. The experimental data are normalized around the weak excitation density conditions, and the dotted line represents the linear behavior normalized around the weak excitation density conditions.

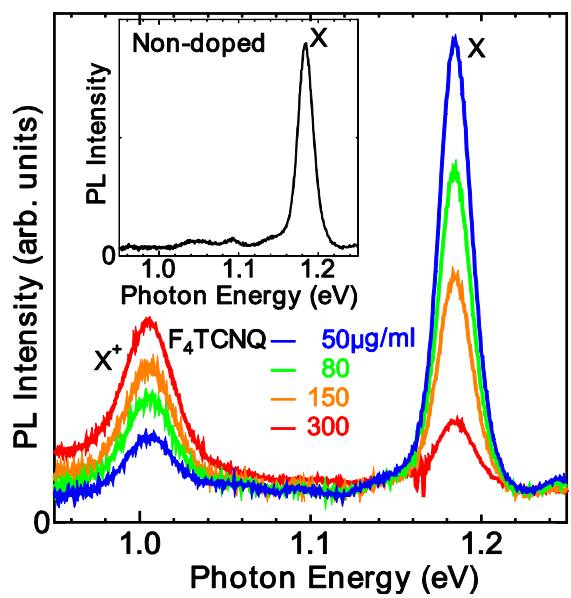

Fig.1    N. Akizuki *et al*.

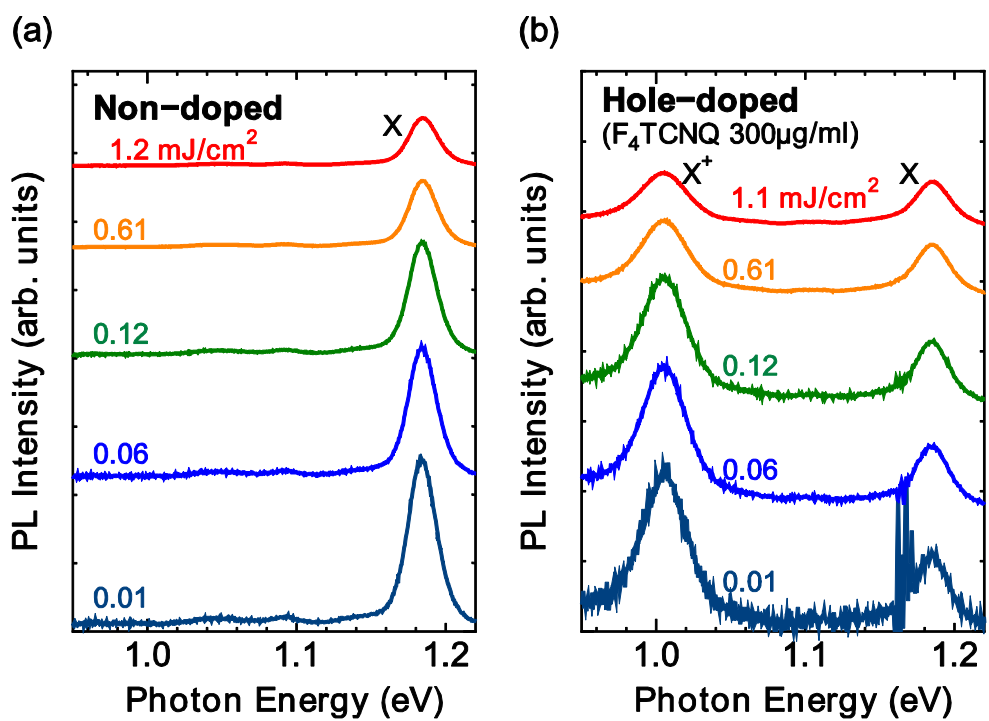

Fig.2  N. Akizuki *et al*.

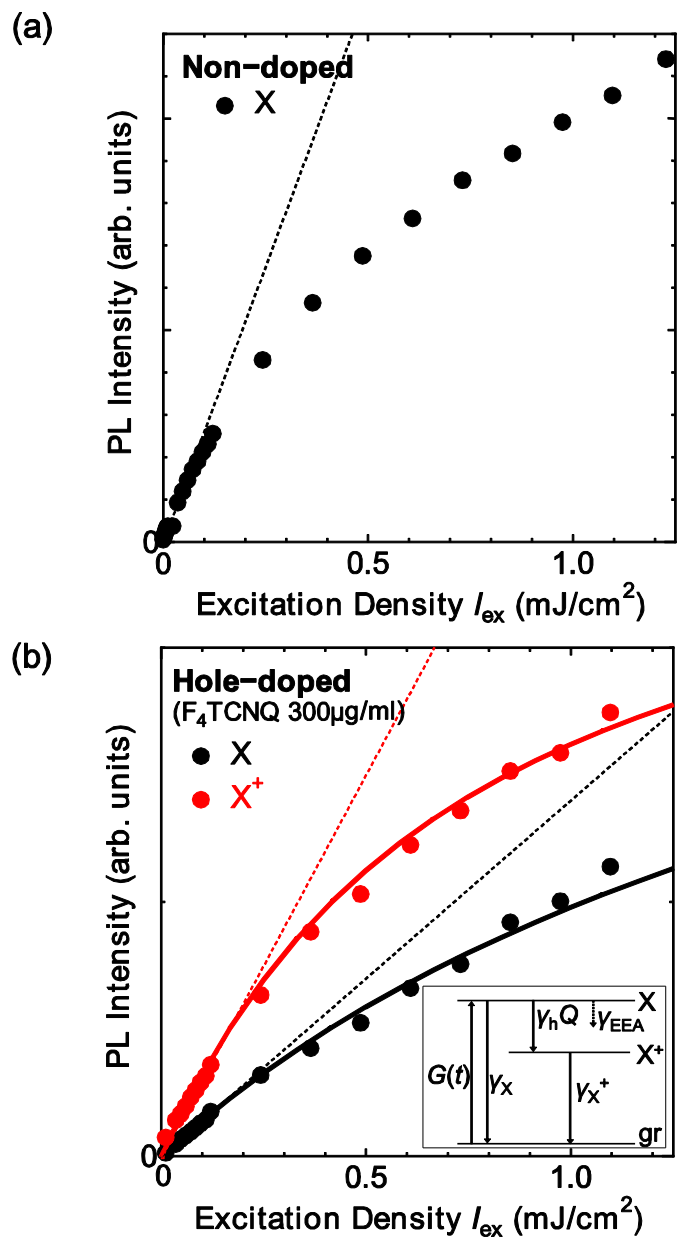

Fig.3  N. Akizuki *et al*.

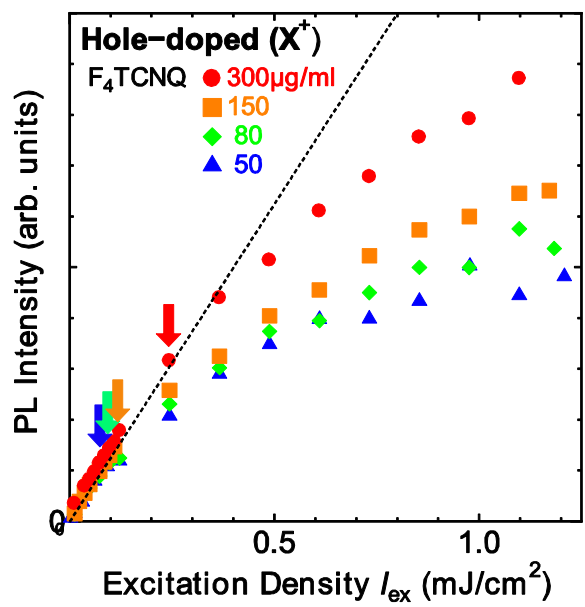

Fig.4    N. Akizuki et al.

# Supplemental Material

# Nonlinear Photoluminescence Properties of Trions in Hole-doped Single-walled Carbon Nanotubes

Naoto Akizuki[1], Munechiyo Iwamura[1], Shinichiro Mouri[1], Yuhei Miyauchi[1,2], Tomohiro Kawasaki[3], Hiroshi Watanabe[3], Tohru Suemoto[3], Kouta Watanabe[4], Kenichi Asano[4], and Kazunari Matsuda[1,+]

[1]*Institute of Advanced Energy, Kyoto University, Uji, Kyoto 611-0011, Japan*

[2]*Japan Science and Technology Agency, PRESTO, 4-1-8 Honcho Kawaguchi, Saitama 332-0012, Japan*

[3]*Institute of Solid State Physics, University of Tokyo, 5-1-5 Kashiwanoha, Kashiwa-shi, Chiba 277-8581, Japan*

[4]*Department of Physics, Osaka University, Toyonaka, Osaka 560-0043, Japan*

## 1. Photoluminescence excitation spectra of non-doped and hole-doped SWNTs

We assumed the relaxation process from higher to lower state of excitons and trions to evaluate the PL quantum yields. This assumption is confirmed as follows. The PL excitation (PLE) spectra were obtained using a supercontinuum light source (SC-400-4, Fianium) with various excitation photon energies from 1.11 to 2.18 eV. Figure S1(a) shows the PLE spectrum probed at the $E_{11}$ exciton peak (1.19 eV, blue circles) and absorption spectrum (solid line) of non-doped (7,5) SWNTs. These spectra are normalized by the intensity of the $E_{11}$ exciton peak (1.19 eV). Figure S1(b) shows the PLE spectra probed at the $E_{11}$ exciton peak (1.19 eV, blue asterisks) and trion ($X^+$) peak (1.01 eV, red circles) of hole-doped (7,5) SWNTs. The absorption spectrum (solid line) of hole-doped (7,5) SWNTs is also shown in Figure S1(b). The PLE and absorption spectra are normalized at the intensity of the $E_{11}$ exciton peak (1.19 eV). The PLE intensities probed at the $E_{ii}$ state, $I_X(E_{ii})$, are expressed as

$$I_X(E_{11}) \propto \eta_X \cdot N_X(E_{11}), \qquad (S1)$$

$$I_X(E_{22}) \propto \eta_X \cdot p_X \cdot N_X(E_{22}), \qquad (S2)$$

where $\eta_X$ is the PL quantum yield of the exciton (X) from the $E_{11}$ to ground state, $p_X$ is the relaxation quantum yield of the exciton (X) from the $E_{22}$ to $E_{11}$ state, and $N_X(E_{ii})$ is the number of excitons (X) in the $E_{ii}$ state. The absorption intensity of the $E_{ii}$ state, $\text{Abs}(E_{ii})$, is expressed as

$$\text{Abs}(E_{11}) \propto N_X(E_{11}), \qquad (S3)$$

$$\text{Abs}(E_{22}) \propto N_X(E_{22}). \qquad (S4)$$

Using Eqs. S1–S4, the relaxation quantum yield of the exciton (X) from the $E_{22}$ to $E_{11}$ state, $p_X$, was evaluated from $I_X(E_{22}) \cdot \text{Abs}(E_{11}) / I_X(E_{11}) \cdot \text{Abs}(E_{22})$. The relaxation quantum yields of the exciton (X) from the $E_{22}$ to $E_{11}$ state, $p_X$ of non-doped and hole-doped SWNTs were estimated to be $1.0 \pm 0.15$ and $0.8 \pm 0.1$, respectively, from the PLE measurements shown in Figs. S1(a) and S1(b). These results suggest that the relaxation probability of the exciton from the higher state to the $E_{11}$ state is nearly unity with negligible losses due to radiative or non-radiative recombination during this relaxation process.

The formation quantum yield of the trion (X$^+$) from the exciton, $p_{X^+}$, in hole-doped SWNTs was also evaluated to be 0.8 ± 0.1 from $I_{X^+}(E_{22}) \cdot \text{Abs}(E_{11})/I_{X^+}(E_{11}) \cdot \text{Abs}(E_{22})$, as shown in Fig. S1(b). This result suggests that the formation probability of a trion from the $E_{11}$ exciton state is also very close to unity. The nearly unity formation probability means that the formation rate of a trion from an exciton is very high in comparison with that of the other processes, which is also confirmed by the analysis of the nonlinear saturation behavior of the trion PL intensity based on the rate equation model in the main text.

## 2. PL behavior of excitons and trions excited by continuous wave and femtosecond pulsed lasers

We compared the PL behavior of excitons and trions using continuous wave (CW) and femtosecond pulsed lasers to understand the physical origin of the observed nonlinear PL behavior. Figure S2 shows the comparison of the excitation density dependence of the trion (X$^+$) PL intensity excited by the CW laser (solid circles) with that by the pulsed laser (open triangles). The strong nonlinear saturation behavior of the PL intensity is only observed under pulsed laser excitation conditions, which suggests that this strong nonlinear PL behavior does not come from an increase in sample temperature but from optically excited excitons and trions.

## 3. Excitation density and the number of excitons

To compare the experimental data given in Fig. 3 with the calculated results of the rate equations, we used the relationship between the number of generated excitons $G = \int_0^\infty G(t)dt$ and the excitation density $I_\text{ex}$, $G = \sigma(h\nu)LI_\text{ex}P_\text{R}/h\nu$, where the absorption cross section $\sigma(h\nu)$ = 25 nm$^2$/μm at the excitation photon energy of 1.55 eV, the average nanotube length $L$ is ~0.3 μm [1], and the randomness of tube directions in the liquid sample $P_\text{R}$ = 0.4. The absorption cross section used here is consistent with previously reported values [2].

## 4. PL saturation behavior of excitons with various doped hole concentrations

Figure S3 shows the excitation density dependence of the integrated PL intensity of the $E_{11}$ exciton (X) in hole-doped SWNTs with various F$_4$TCNQ concentrations. The experimental curves in Fig. S3 are normalized around the weak excitation density conditions, and the dotted line shows the linear behavior. The nonlinear behaviors of the integrated PL intensities of the $E_{11}$ exciton (X) peak in hole-doped SWNTs are equivalent below the F$_4$TCNQ concentration of 150 μg/ml, where the average distance of doped holes is $Q^{-1}$~14 nm. As shown in Fig. S3, the PL saturation at lower concentrations below 150 μg/ml shows same behavior, which is stronger than that at the F$_4$TCNQ concentration of 300 μg/ml ($Q^{-1}$~6 nm). This experimental result can be explained by the balance between the characteristic length scale of average doped hole distance ($Q^{-1}$) and exciton coherence (~10 nm) [3]. The average distance of doped holes in heavily hole-doped SWNTs is shorter than the exciton coherence length, and the exciton relaxation to trion states predominantly occurs relative to EEA processes. Moreover, the EEA processes themselves are restrained because the two excitons are hardly generated simultaneously within a length divided by doped holes, which reduces the EEA rate in SWNTs.

## 5. PL decay of excitons and trions in hole-doped SWNTs

We measured the PL decay of excitons and trions in SWNTs hole-doped with a F$_4$TCNQ concentration of 150 μg/ml by the up-conversion method. A Ti:Al$_2$O$_3$ pulsed laser (Coherent, Mira Seed) amplified by a regenerative amplifier (Coherent, RegA9050) with an excitation photon energy of 1.57 eV, a repetition rate of 200 kHz, and pulse width of 70 fs was divided into gate and pump beams, and the sum frequency was detected by a photomultiplier tube (Hamamatsu Photonics, R943-02). Figure S4 shows the PL decay dynamics probed around the $E_{11}$ exciton (X) peak (1.2 eV, black circles) and trion (X$^+$) peak (1.0 eV, red circles) of hole-doped SWNTs in the weak excitation condition of 0.016 mJ/cm$^2$ where EEA processes do not occur. The rate equations under weak excitation conditions are given as

$$\frac{dN_X(t)}{dt} = G(t) - \gamma_X N_X(t) - \gamma_h Q N_X(t), \quad (S5)$$

$$\frac{dN_{X^+}(t)}{dt} = \gamma_h Q N_X(t) - \gamma_{X^+} N_{X^+}(t). \quad (S6)$$

The experimentally obtained decay curves of exciton (X) and trion (X$^+$) peaks are well reproduced by the calculated curves in Fig. 4 by solving the above rate equations. Thus, the PL decay time probed around the $E_{11}$ exciton (X) peak (~400 fs) comes from the trion formation time, and the obtained value is consistent with the value (450 fs) calculated from the rate equation models in the main text. The obtained PL lifetime probed around the trion (X$^+$) peak (~580 fs) is nearly consistent with the calculated value (~800 fs)$^{-1}$ from the rate equations in the main text and the reported value [4].

(a)
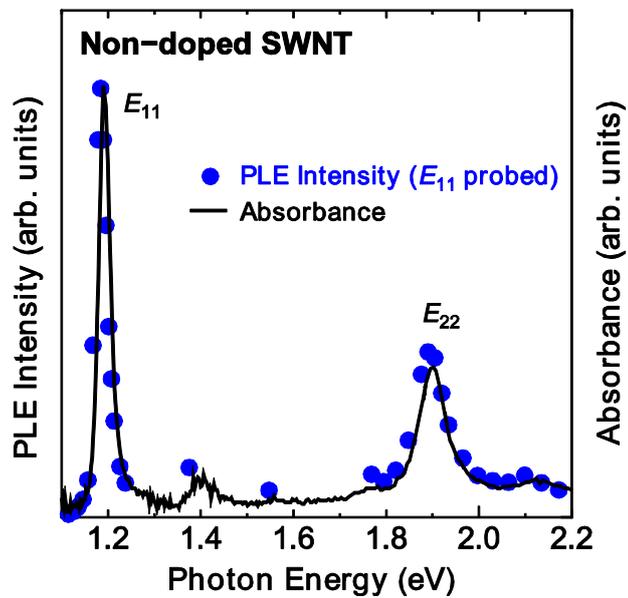

(b)
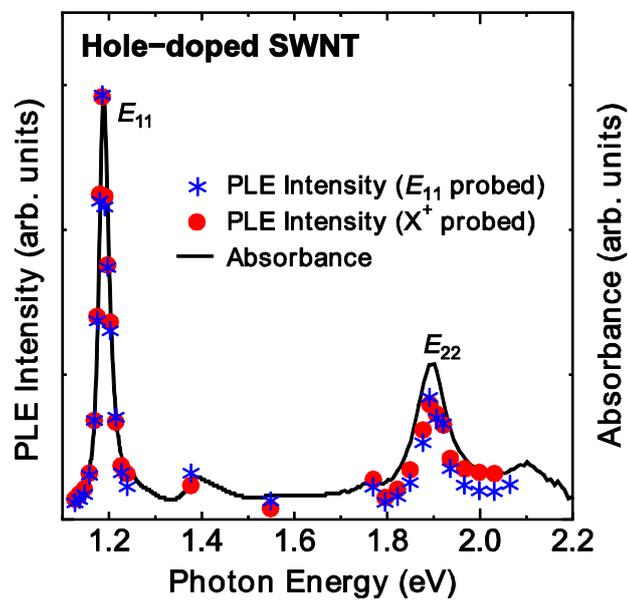

Fig.S1    N. Akizuki *et al.*

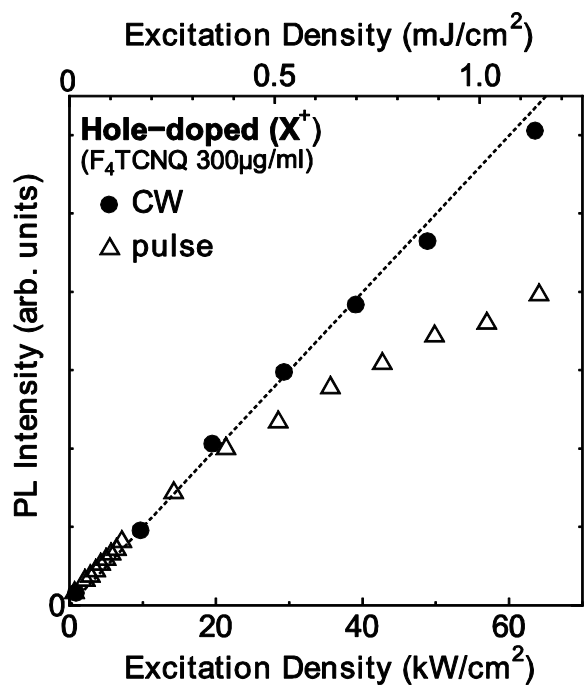

Fig.S2    N. Akizuki *et al*.

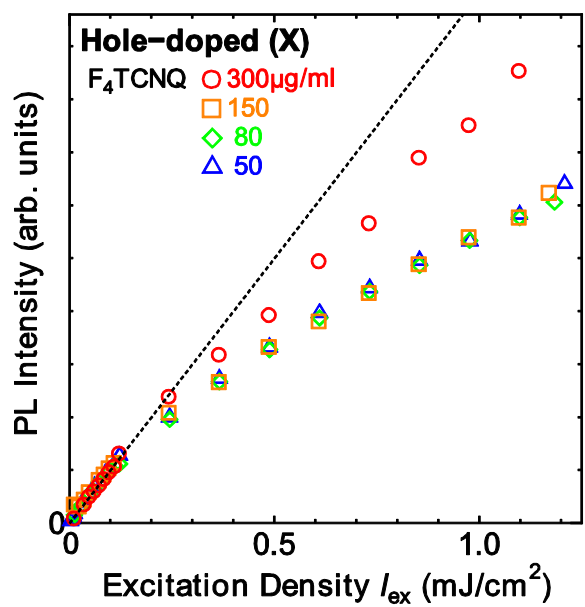

Fig.S3    N. Akizuki *et al*.

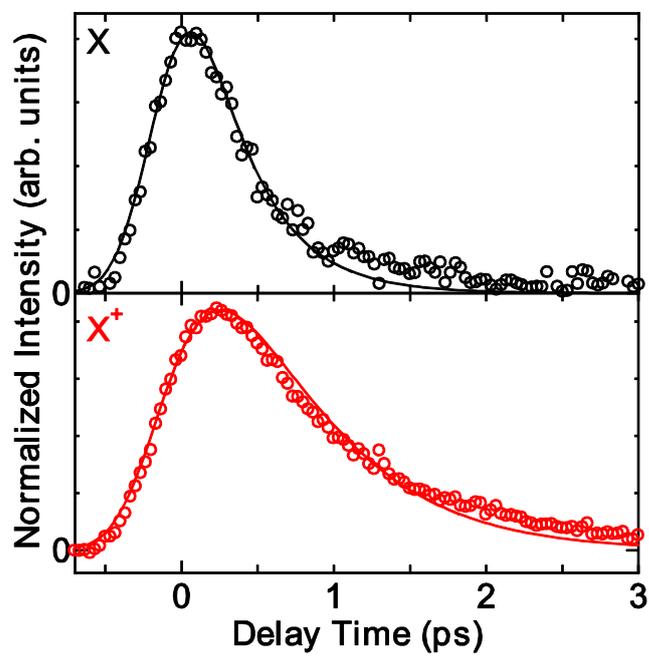

Fig.S4    N. Akizuki *et al*.

29